\documentclass[aps,prl,twocolumn,showpacs,superscriptaddress]{revtex4}
\usepackage{graphicx} % Include figure files
\def\Journal#1#2#3#4{{#1} {\bf #2}, #3 (#4)}
\def\NIMA{{\rm Nucl. Instr. Meth.} A}
\def\NPA{{\rm Nucl. Phys.} A}
\def\PRL{\rm Phys. Rev. Lett.}
\def\PRC{{\rm Phys. Rev.} C}
\def\FBS{\rm Few Body Syst} 
\def\PL{\rm Phys. Lett.}
\def\ARNPS{\rm Ann. Rev. Nucl. Part. Sci.}
\begin{document}

\title{Evidence for High-Momentum Enhancement in the Exclusive
$^{3}\!He(e,e^{\prime}p)$ Reaction Below the
Quasi-Elastic Peak}

\author{A.~Kozlov}
\affiliation{Department of Physics, University of Regina, Regina,
              SK S4S0A2, Canada}
\author{A.J.~Sarty}
\affiliation{Department of Astronomy \& Physics, Saint Mary's University,
              NS B3H3C3, Canada}
\author{K.A.~Aniol}
\affiliation{Department of Physics and Astronomy, California State
              University, Los Angeles, CA 90032, USA}
\author{P.~Bartsch}
\author{D.~Baumann}
\affiliation{Institut f\"ur Kernphysik, Universit\"at Mainz,
              D-55099 Mainz, Germany}
\author{W.~Bertozzi}
\affiliation{Laboratory for Nuclear Science, 
              MIT, Cambridge, MA 02139, USA}
\author{K.~Bohinc}
\affiliation{Institute ``Jo\v zef Stefan'', 
              University of Ljubljana, SI-1001 Ljubljana, Slovenia}
\affiliation{Institut f\"ur Kernphysik, Universit\"at Mainz,
              D-55099 Mainz, Germany}
\author{R.~B\"{o}hm}
\affiliation{Institut f\"ur Kernphysik, Universit\"at Mainz,
              D-55099 Mainz, Germany}
\author{J.P.~Chen}
\affiliation{Thomas Jefferson National Accelerator Facility,
	      Newport News, VA 23606, USA} 
\author{D.~Dale}
\affiliation{Department of Physics and Astronomy,
              University of Kentucky, Lexington, KY 40506, USA}
\author{L.~Dennis}
\affiliation{Department of Physics,
              Florida State University, Tallahassee, FL 32306, USA}
\author{S.~Derber}
\author{M.~Ding}
\author{M.O.~Distler}
\affiliation{Institut f\"ur Kernphysik, Universit\"at Mainz,
              D-55099 Mainz, Germany}
\author{P.~Dragovitsch}
\affiliation{Department of Physics,
              Florida State University, Tallahassee, FL 32306, USA}
\author{I.~Ewald}
\author{K.G.~Fissum}
\affiliation{Laboratory for Nuclear Science, 
              MIT, Cambridge, MA 02139, USA}
\author{J.~Friedrich}
\author{J.M.~Friedrich}
\author{R.~Geiges}
\affiliation{Institut f\"ur Kernphysik, Universit\"at Mainz,
              D-55099 Mainz, Germany}
\author{S.~Gilad}
\affiliation{Laboratory for Nuclear Science, 
              MIT, Cambridge, MA 02139, USA}
\author{P.~Jennewein}
\author{M.~Kahrau}
\affiliation{Institut f\"ur Kernphysik, Universit\"at Mainz,
              D-55099 Mainz, Germany}
\author{M.~Kohl}
\affiliation{ Institut f\"ur Kernphysik, Technische Universit\"at
              Darmstadt, D-64289 Darmstadt, Germany}
\author{K.W.~Krygier}
\author{A.~Liesenfeld}
\affiliation{Institut f\"ur Kernphysik, Universit\"at Mainz,
              D-55099 Mainz, Germany}
\author{D.J.~Margaziotis}
\affiliation{Department of Physics and Astronomy, California State
              University, Los Angeles, CA 90032, USA}
\author{H.~Merkel}
\author{P.~Merle}
\author{U.~M\"{u}ller}
\author{R.~Neuhausen}
\author{T.~Pospischil}
\affiliation{Institut f\"ur Kernphysik, Universit\"at Mainz,
              D-55099 Mainz, Germany}
\author{M.~Potokar}
\affiliation{Institute ``Jo\v zef Stefan'', 
              University of Ljubljana, SI-1001 Ljubljana, Slovenia}
\author{G.~Riccardi}
\author{R.~Roch\'{e}}
\affiliation{Department of Physics,
              Florida State University, Tallahassee, FL 32306, USA}
\author{G.~Rosner}
\affiliation{Institut f\"ur Kernphysik, Universit\"at Mainz,
              D-55099 Mainz, Germany}
\author{D.~Rowntree}
\affiliation{Laboratory for Nuclear Science, 
              MIT, Cambridge, MA 02139, USA}
\author{H.~Schmieden}
\affiliation{Institut f\"ur Kernphysik, Universit\"at Mainz,
              D-55099 Mainz, Germany}
\author{S.~ \v Sirca}
\affiliation{Institute ``Jo\v zef Stefan'', 
              University of Ljubljana, SI-1001 Ljubljana, Slovenia}
\author{J.A.~Templon}
\affiliation{Department of Physics and Astronomy,
              University of Georgia, Athens, GA 30602, USA}
\author{M.N.~Thompson}
\affiliation{School of Physics, The University of Melbourne, 
              VIC 3010, Australia}
\author{A.~Wagner}
\author{Th.~Walcher}
\author{M.~Weis}
\affiliation{Institut f\"ur Kernphysik, Universit\"at Mainz,
              D-55099 Mainz, Germany}
\author{J.~Zhao}
\author{Z.-L.~Zhou}
\affiliation{Laboratory for Nuclear Science, 
              MIT, Cambridge, MA 02139, USA}
\collaboration{A1 Collaboration}\noaffiliation
\date{\today}

\begin{abstract}
New, high-precision measurements of the
$^{3}\!He(e,e^{\prime}p)$ reaction
using the A1 collaboration spectrometers at the Mainz microtron MAMI 
are presented. These were performed in parallel kinematics
at energy transfers below the quasi-elastic peak, and at a
central momentum transfer of 685 MeV/c.  Cross sections and
distorted momentum distributions were extracted
and compared to theoretical predictions and existing data. The
longitudinal and transverse behavior of the cross section was also studied.
Sizable enhancements of the cross
sections for missing momenta larger than 100 MeV/c as compared to Plane Wave
Impulse Approximation were observed in both the two- and three-body breakup
channels.
\end{abstract}

% see http://www.aip.org/pacs/pacs99/

\maketitle

% body of paper here

Renewed precision studies of few-nucleon systems has been fueled
by recent developments on both the theoretical and experimental fronts.
Theoretically, progress toward full microscopic calculations based on realistic
$NN$ potential models has been achieved - for example, via nonrelativistic
Faddeev-type calculations for three-body systems \cite{Golak:95, Kotlyar:00},
and via Monte Carlo variational calculations for three- and four-body systems
\cite{Carlson:94}.  Experimentally, new facilities such as the Mainz microtron
MAMI and the Thomas Jefferson National Accelerator Facility provide access 
to high-quality continuous-wave electron beams,
high-resolution magnetic spectrometers 
and fast data-acquisition systems. These features allow precision measurements
of the nuclear electromagnetic response through coincidence proton-knockout 
electron scattering. Thus, the door is opened for detailed studies of 
unresolved issues relating to few-nucleon systems,
some of which were first raised in inclusive
$(e,e^{\prime})$ measurements more than 20 years ago.

The experiment reported in this paper is part
of an extensive and systematic program~\cite{proposal} 
to study nucleon knockout reactions on $^{3,4}\!He$ nuclei
over the quasi-elastic (QE) peak at a fixed central
momentum-transfer of $| \vec{q} |$ = 685 MeV/c.
  The specific measurements reported here were performed
with $^{3}\!He$ on the low-energy side of the QE peak
(fixed central energy-transfer, $\omega \approx$ 158 MeV;
$x_B \approx 1.5$),
where the contributions from two-body mechanisms
such as meson-exchange currents (MEC) and isobar currents (IC) are 
suppressed, thereby 
enhancing the possibility of observing the effects of short-range $NN$
correlations.
%
% First Figure (called "fig:xs")
%
\begin{figure}[t]
\includegraphics{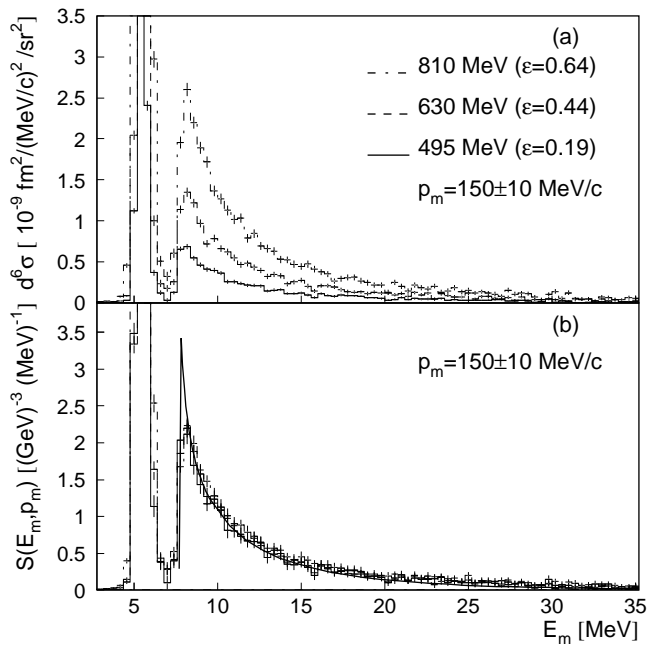}
\caption{\label{fig:xs} (a) Radiatively-corrected 
six-fold $^{3}\!He(e,e'p)X$ cross sections shown with 
the two-body breakup peak positioned at 5.49~MeV, and the continuum
channel threshold at 7.72~MeV.
(b) The experimental spectral function for the three measured
$\epsilon$-values compared to the theoretical calculation
from \cite{Kievsky:97}. At $E_m \ge$30~MeV, the measured spectral
function was close to zero.}
\end{figure}
This low energy-transfer region was studied extensively in the 1970-80's
using inclusive electron scattering at $| \vec{q} | \gg k_F$
\cite{McCarthy:76, Day:78, Sick:80}.
The measured $^{3}\!He(e,e')$ cross
sections in this kinematical regime were much higher than theoretical
impulse approximation (PWIA)
predictions
\cite{Sick:80, degliAtti:83}, and  
the discrepancy with prediction increased as $\omega$ decreased (i.e. as
$x_B$ increased).  Since these inclusive cross sections were measured over
a wide range of momentum transfer and exhibited so-called ``$y$-scaling"
in this regime, it has been argued that the 
discrepancies cannot be accounted for by non-quasifree mechanisms
such as MEC or IC \cite{Day:90}.  Further, effects of final-state interactions
(FSI) in this region were estimated to be small \cite{Sauer:89}.
It was suggested \cite{Sick:80} that,
as a possible solution to this discrepancy,
the high-momentum components in the $^{3}\!He$ spectral function should
be increased.  Such an increase would be indicative of short-range NN
correlations not accounted for in the theory.
This suggestion came under
scrutiny, and it was subsequently suggested \cite{degliAtti:83} that 
exclusive and semi-exclusive
$^{3}\!He(e,e^{\prime}p)$ measurements were required
to test such a modification, and to also examine the missing-energy ($E_m$)
dependence of the spectral function.  Thus, a later measurement of
$^{3}\!He(e,e'p)$ at high missing momenta was performed
\cite{Marchand:88}.  It was reported that the new data did not support
{\it ad hoc} attempts to
increase the spectral function high-momentum components - 
leading to the conclusion that the observed enhanced strength in 
the earlier inclusive data (as discussed in Ref. \cite{Sick:80})
was more likely due to defects in
the PWIA model assumptions, than to a lack of nucleon high-momentum 
components in the $^{3}\!He$ wave function.  However, it should be
noted that this exclusive data of Ref. \cite{Marchand:88} sampled
the kinematical domain with $\omega$ above the QE peak, in the ``dip"
region, and not the low-energy side at which the inclusive data are
discussed in Ref. \cite{Sick:80}.  It had been in the low-energy region
where $y$-scaling was observed, indicating that single-nucleon reaction
mechanisms dominate.  It can also be noted that none of the
since-reported exclusive $^{3}\!He(e,e'p)$ measurements
\cite{Jans:87, Ducret:93, LeGoff:97, Florizone:99} have been carried out
in the kinematical conditions studied by Ref.~\cite{Sick:80}.
In this work, we report the first
$^{3}\!He(e,e'p)$ measurements on the low-energy
side of the QE peak.  These new exclusive data were taken in
``parallel kinematics" (central
detected-proton angle equal to central angle of $\vec{q}$), and
show an enhancement
of strength in the high-momentum region of the spectral function.  Due to
the parallel nature of the measurement, this is directly related to the
$\omega$-dependence.  Therefore, the enhancement shows up on the low-energy
side of the QE peak, similar to what was previously observed for the
inclusive reaction.

The measurements utilized
three incident beam energies $E_{i}$~=~495, 630, and 810~MeV
with electron scattering angles $\Theta_e$~= 109.28$^\circ$, 
75.31$^\circ$, and 54.58$^\circ$, corresponding to virtual
photon polarizations of $\epsilon$~= 0.19, 0.44, and 0.64.
The total systematic uncertainty of the measured cross sections
was estimated to be 
$\pm3$\%. This uncertainty is dominated by the uncertainty in the
density of the $^{3}He$ gas in the cold ($T$ = 20 K), high-pressure
($P$ = 18 atm) target.
All error bars shown in the figures of this paper 
are statistical only. 
More details on the experimental setup, performance, and data-analysis 
procedure can be found in Ref.\cite{Blomqvist:98} and \cite{Kozlov:99}. 

The $(e,e'p)$ cross section for $^{3}\!He$ was obtained as a
function of missing energy and missing momentum ($p_m$).
The radiative unfolding procedure
using the Mainz version of RADCOR code \cite{Rokavec:94} was applied to
correct for radiation in 2-D ($E_m, p_m$) space.
The extracted $(e,e'p)$ cross section
gradually approached zero at
$E_m$~=~30~MeV. An alternative method was
used to test this radiative unfolding procedure: radiative effects were 
included directly into 
a MC simulation code (``MCEEP" \cite{mceep}), following the method outlined
in Ref.~\cite{Templon:00}.  This method confirmed that most
of the measured $(e,e'p)$ strength beyond $E_m$~=~30~MeV arises
from the processes at lower $E_m$, and that the actual (unradiated)
strength at these high values of$E_m$ was too small to be extracted.

Distorted spectral functions, $S^{dist}(E_{m},p_{m})$, were
extracted from the data according to:
\begin{equation} 
S^{dist}(E_{m},p_{m})=\frac{1}{p^{2}_{p}\sigma_{ep}^{cc1}}
\frac{d^{6}\sigma}{d\Omega_{e}d\Omega_{p} dp_{e}dp_{p}} ~.
\end{equation}
%
% Second Figure (called "fig:2bbu" in text)
%
\begin{figure}[t]
\includegraphics{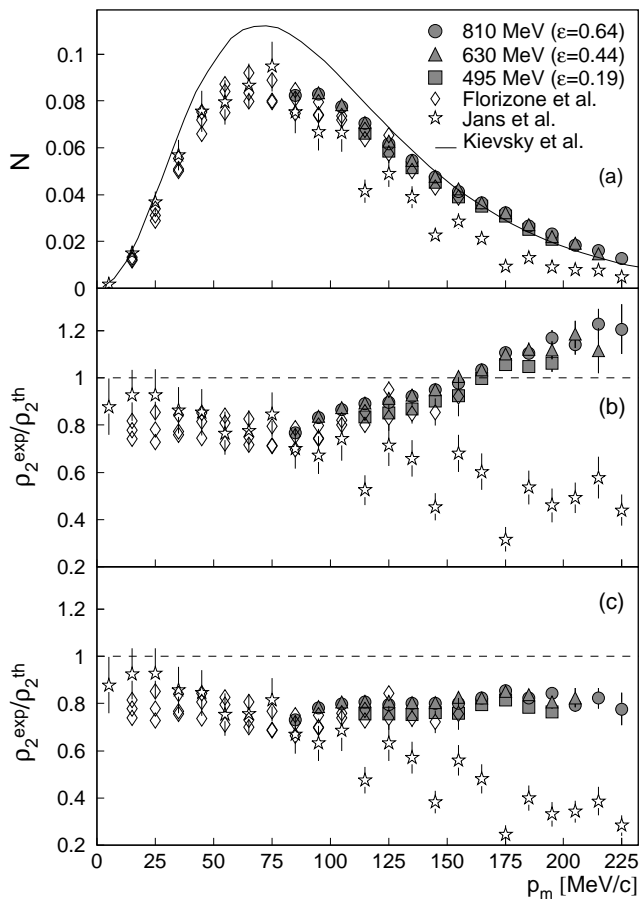}
 \caption{\label{fig:2bbu}
      Distorted proton-momentum distributions, $\rho_2$,
      for the $^{3}\!He(e,e'p)d$ two-body breakup channel.  Extracted
      measurements from the three values of $\epsilon$ are shown as
      filled symbols.  Earlier data from MAMI ~\cite{Florizone:99}
      and Saclay ~\cite{Jans:87} are shown as open symbols.
      Note the Saclay data of Ref.~\cite{Jans:87} were taken under different
      kinematical conditions than the other data, being non-parallel and fully
      centered at $x_B$ = 1.
      {\it (a)} Measured $\rho_2$ values multiplied by
      $4 \pi p_m^2 \Delta_{p_m}$: $N = 4 \pi p_m^2 \rho_2 \Delta_{p_m}$,
      with $\Delta_{p_m}$ equal to the bin-width of 10 MeV/c.  The solid
      line is the spectral-function calculation of
      Kievsky {\em et al.} ~\cite{Kievsky:97}.  {\it (b)}
      Ratio of extracted proton-momentum distributions ($\rho_2^{exp}$)
      to the spectral function prediction ($\rho_2^{th}$) of Kievsky
      {\em et al.} \cite{Kievsky:97}.  {\it (c)} Same ratio as in plot
      {\it (b)}, except now obtaining $\rho_2^{th}$ by multiplying the
      spectral function of Ref. ~\cite{Kievsky:97} by the renormalization
      factor $f(k)$ as suggested in Ref. ~\cite{Sick:80}.
      }
\end{figure}
Study of the 
longitudinal/transverse ($L/T$) behavior of the cross section was done by
direct comparison to the $e-p$ off-shell cross section on the moving
proton, at
each $\epsilon$, since pure PWIA quasi-free knockout would follow the
elementary $e-p$ response.
We chose the de Forest $cc1$ prescription
\cite{Forest:83} for the elementary $e-p$ cross section
($\sigma_{ep}^{cc1}$), mainly because of
convenience to compare our results to earlier data from Ref.
\cite{Florizone:99} and Ref. \cite{Jans:87}.
A further benefit to dividing out of the
cross section some 
well-defined form of $\sigma_{eN}$ is that the predominant
kinematic ($\epsilon$) dependence is removed from the data, enabling 
easier identification of residual nuclear dependencies. 

The measured six-fold cross section for the reaction $^{3}\!He(e,e'p)X$
as a function of $E_m$, averaged over 
one sample $p_m$ bin of 150$\pm10$~MeV/c,
is shown for the three different $\epsilon$ values in Fig.~\ref{fig:xs}(a).
The two-body
$^{3}\!He(e,e'p)d$ breakup peak and the threshold for three-body
$^{3}\!He(e,e'p)pn$ breakup are at
5.49~MeV and 7.72~MeV respectively.
The strong dependence of the
cross section on the virtual-photon polarization is evident.
This dependence, however, disappears by dividing out the elementary
$ep$ cross section in the extraction of the
measured spectral function, as can be seen from Fig.~\ref{fig:xs}(b).
The residual $\epsilon$-dependence is
less than 5\% between the differing $\epsilon$ measurements, and
has no systematic trend,
meaning that the $L/T$ behavior of the $(e,e'p)$ cross section is fully
described by the $\sigma_{ep}$ cross section.

Two-body $p-d$ distorted momentum distributions, $\rho_2$,
were obtained by
integrating the extracted $S^{dist}$ over the two-body breakup peak.  The
extracted $\rho_2$ distributions are shown in
Fig.~\ref{fig:2bbu}(a), which shows $\rho_2$ in terms of
$N = 4 \pi p_m^2 \rho_2 \Delta_{p_m}$, where $\Delta_{p_m}$ is
the bin-size in $p_m$ (equal to 10 MeV/c here).
Also shown in Fig.~\ref{fig:2bbu}(a) are earlier data
\cite{Florizone:99, Jans:87} and the spectral function calculation of
Kievsky {\em et al.} \cite{Kievsky:97}.
Like the spectral function, our data are independent of $\epsilon$ for
the entire $p_m$ range.

By plotting the data as a ratio with respect to
the Kievsky spectral function, as is done in Fig.~\ref{fig:2bbu}(b), several
points are notable.
First, the data of Ref.~\cite{Jans:87} -- which was taken on top of the QE
peak, but is the only data set taken in transverse kinematics,
thus at constant $\omega$ -- display reasonable agreement with those
of Ref.~\cite{Florizone:99} for $p_m$ up to 95 MeV/c, but deviate both from
our new data and those of Ref.~\cite{Florizone:99} at higher $p_m$.
Second,
agreement is seen between the current measurement and the earlier parallel
kinematics Mainz
data of Ref.~\cite{Florizone:99}, which was taken at the top of the QE peak.
Third, a roughly constant 20\% suppression of the data 
compared to the spectral function is seen for $p_m$ up to 100 MeV/c.
This roughly 20\% difference between data and the calculation at low $p_m$
has been previously observed (e.g. see Ref.~\cite{Florizone:99}) and results 
predominantly from the fact that the theoretical spectral function does not 
account for FSI while the data does.

The notable new feature
observed in Fig.~\ref{fig:2bbu}(b) is the significant relative 
increase in the measured cross section
compared to the spectral function for $p_m$ above 100 MeV/c.
The beginning of this phenomenon of relative increase
can be also seen in the few highest $p_m$ points of Ref.~\cite{Florizone:99}.
It should be noted that the parallel kinematics data 
from Ref.~\cite{Florizone:99} extend from the top of the QE peak to the
low-energy side, and that, just as for our new data, higher $p_m$
corresponds to lower $\omega$.  Similar decrease is not seen in
Ref.~\cite{Jans:87}, which in contrast to our new
data, show a steady decrease for increasing $p_m$
with respect to the calculation.
This suggests that the (relative) increase in the
cross section is a phenomenon observed only at the
low-energy side of the QE peak.  Moreover, our
observed enhancement in strength appears to be consistent with that 
observed in the earlier inclusive measurements ~\cite{McCarthy:76, Day:78,
Sick:80}. This can be demonstrated by multiplying
the spectral function of Kievsky {\em et al.} by the enhancement factor that
was suggested in Ref.~\cite{Sick:80}:
$f(k)=1+(k/\tilde{k})^n$
(where $n$=2.5, $\tilde{k}$=285~MeV/c, and $k$ is the primordial proton 
momentum in the nucleus, using $k$ = $p_m$ here).  When the theory is
modified in this way, the $\rho_2^{exp}/\rho_2^{th}$ ratio 
for our new data
becomes mainly constant at roughly 0.8 over the entire $p_m$ range up to
225 MeV/c.  This is demonstrated in Fig.~\ref{fig:2bbu}(c).  Another possible
explanation for 
the enhancement feature being manifest only at $x_B ~>~1$ may be a
non-quasifree reaction mechanism, such as a
a QE reaction on a quasi-deuteron
with subsequent rescattering; this would represent a longer-range nuclear
correlation effect.
%
% Third figure (called "fig:3bbu" in text)
%
\begin{figure}[t]
\includegraphics{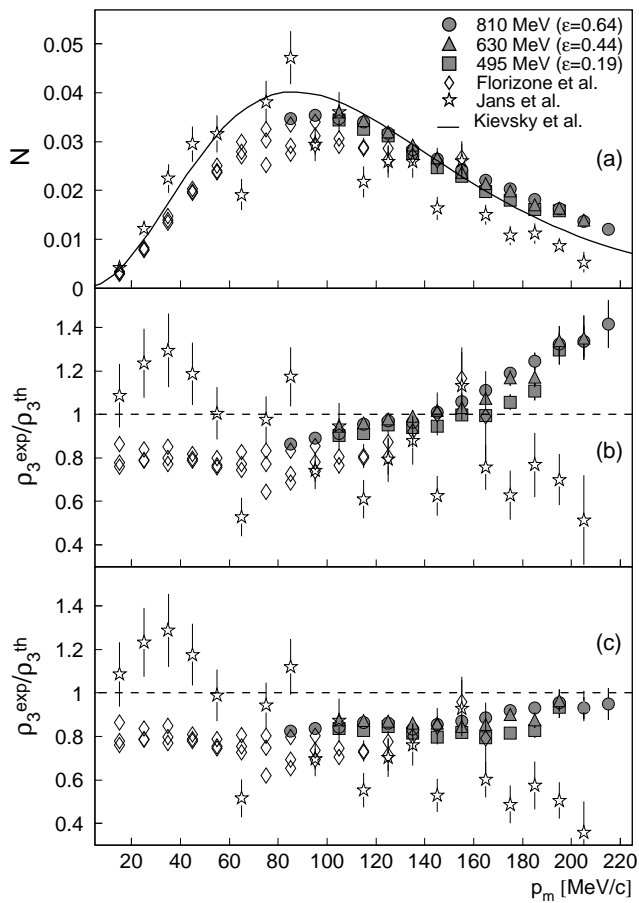}
 \caption{\label{fig:3bbu}
      Distorted proton-momentum distributions, $\rho_3$,
      for the $^{3}\!He(e,e'p)pn$ three-body breakup channel.  Plots and
      symbols follow the identical format used for
      $\rho_2$ in Fig.~\ref{fig:2bbu}.
      }
\end{figure}

Distorted proton-momentum distributions for the three-body
breakup channel, $\rho_3$, were obtained by integration of the
experimental spectral function $S^{dist}(E_{m},p_{m})$ from 7 to
20~MeV in $E_m$.  These extracted $\rho_3$ distributions are shown in
Fig.~\ref{fig:3bbu}(a), mutliplied by $4 \pi p_m^2 \Delta_{p_m}$ as
was done for the two-body channel.
Again, no systematic $\epsilon$-dependence is observed.  The
ratio $\rho_3^{exp}$/$\rho_3^{th}$ is shown as a function of $p_m$ in
Figs.~\ref{fig:3bbu}(b) and \ref{fig:3bbu}(c) using the unaltered and the
modified calculation of
Kievsky {\em et al.}, respectively.  These ratios for the continuum
channel show a similar feature as seen in the two-body channel.
The modification of the spectral function by the factor $f(k)$
leads to a better
agreement with the data, but not as good as was observed for the two-body
breakup channel.
  
In conclusion, we have performed measurements of the $^{3}\!He(e,e'p)$
reaction on the low $\omega$ side of the QE peak.
Cross sections, distorted spectral functions and distorted proton-momentum
distributions were obtained as a function of $E_m$ and 
$p_m$ for both the two-body and continuum breakup channel
for three $\epsilon$ values. The $(e,e'p)$ cross
section strength falls to near zero for $E_m ~>~$30~MeV.
The strong $\epsilon$-dependence of the
cross section, both for two- and three-body reaction channels,
is due to the $L/T$ behavior of the elementary $e-p$ cross section.
The distorted proton-momentum distributions for 
both the two- and three-body breakup channels significantly
deviate in shape from the PWIA predictions of Ref.~\cite{Kievsky:97}
for $p_m$ above 100 MeV/c.  The correction function
suggested in Ref.~\cite{Sick:80} to enhance the high-momentum
components of the spectral function, based on
inclusive measurements in this same kinematical region ($\omega <
\omega_{QE}$), is able to fully correct
this deviation in shape for the two-body breakup channel, and largely
for the three-body breakup channel.
Such an
enhancement of the spectral function's nucleon high-momentum components could
be indicative of the effects of short-range NN correlations.
The remaining difference in amplitude is 
predominantly (though not necessarily exclusively) due to
by FSI effects, which were not included into the calculations. 

We would like to thank the MAMI staff
for providing support for this experiment.
This research was supported by the state of Rhineland-Palatinate and
by grants from the Deutsche Forschungsgemeinschaft (SFB 443),
by the U.S. Department
of Energy and the National Science Foundation, by the Natural Sciences
and Engineering Research Council, and by the University of Melbourne.

\end{document}